# Spatially Resolved High Voltage Kelvin Probe Force Microscopy: A Novel Avenue for Examining Electrical Phenomena at Nanoscale

*Conor J. McCluskey[#], Niyorjyoti Sharma[#], Jesi R. Maguire, Serene Pauly, Andrew Rogers, TJ Lindsay, Kristina M. Holsgrove, Brian J. Rodriguez, Navneet Soin, John Marty Gregg, Raymond G. P. McQuaid, and Amit Kumar\**

E-mail: a.kumar@qub.ac.uk
[#] These authors contributed equally to the work

C. McCluskey, N. Sharma, J. Maguire, S. Pauly, A. Rogers, T. Lindsay, K. M. Holsgrove, J. M. Gregg, R. G. P. McQuaid, A. Kumar
School of Mathematics and Physics, Centre for Quantum Materials and Technologies, Queen's University Belfast, Belfast, BT7 1NN, United Kingdom

N. Soin
School of Engineering, Ulster University, 2-24 York Street, Belfast BT15 1AP, United Kingdom
School of Science, Computing and Engineering Technologies, Swinburne University of Technology, P.O. Box 218, Hawthorn VIC 3122, Australia

B. J. Rodriguez
School of Physics, University College Dublin, Belfield, Dublin 4, Ireland

**Abstract:** Kelvin probe microscopy (KPFM) is a well-established scanning probe technique, used to measure surface potential accurately; it has found extensive use in the study of a range of materials phenomena. In its conventional form, KPFM frustratingly precludes imaging samples or scenarios where large surface potential exists or large surface potential gradients are created outside the typical ±10V window. If the potential regime measurable via KPFM could be expanded, to enable precise and reliable metrology, through a high voltage KPFM (HV-KPFM) adaptation, it could open up pathways towards a range of novel experiments, where the detection limit of regular KPFM has so far prevented the use of the technique. In this work, HV-KPFM has been realised and shown to be capable of measuring large surface potential and potential gradients with accuracy and precision. The technique has been employed to study a range of materials (positive temperature coefficient of resistivity ceramics, charge storage fluoropolymers and pyroelectrics) where accurate spatially resolved mapping of surface potential within high voltage regime facilitates novel physical insight. The results demonstrate that HV-KPFM can be used as an effective tool to fill in existing gaps in surface potential measurements while also opening routes for novel studies in materials physics.

## 1. Introduction

Kelvin Probe Force microscopy (KPFM) is a scanning probe technique which allows quantitative spatially resolved mapping of the Contact Potential Difference (CPD) or the surface potential on the nanoscale.[1] For metals and semiconductors, the measured surface potential is reflective of the local work function differences between the tip and the sample,[2] but, in the case of insulators, the primary contributions to the potential come from the presence of uncompensated surface charges.[3] KPFM continues to be a heavily used scanning probe technique for the study of a range of physical phenomena in functional oxides and two-dimensional materials.[4-6] It is particularly useful in studying ferroelectrics, where changes in surface charge species and densities can provide insight into the static or dynamic properties of the bulk polarised domains.[7,8]

Frustratingly, many current implementations of the KPFM technique preclude imaging large surface potentials exceeding ±10V or in scenarios in which large surface potential gradients exist. This constraint likely arises from the technique's historical emphasis on measuring the work function in metals and semiconductors, where differences exceeding ±10V are improbable, given that the reported work function values for elements range from about 2 to 6 eV. Nevertheless, the extension of KPFM into the high voltage (HV) regime (HV-KPFM) presents an intriguing avenue for uncovering deeper insights into material behavior, particularly within the context of dielectrics and ferroelectrics. These materials can give rise to substantial surface potentials, on the order of hundreds to thousands of volts in bulk crystals and films and such high potentials can lead to novel functionalities. Regrettably, conventional KPFM falls well short in measuring such potentials effectively due to its inherent limitations.

In the context of ferroelectrics and dielectrics, many interesting problems and opportunities have emerged recently where spatially resolved HV-KPFM can provide clinching experimental evidence to help unveil new physical insights. Two clear examples are quantitative Hall voltage mapping[8] and 4-probe characterisation of conducting domain walls in situations where the lateral driving voltage across walls is larger than the typical range achievable through KPFM. In triboelectric dielectrics, large voltages can develop during the charging and contact-electrification processes and HV-KPFM can offer insight through direct visualisation of the spatial distribution of such charging-discharging processes. HV-KPFM also could offer a viable platform for studying the surface potential development in pyroelectric crystals for small temperature changes, which is not possible using conventional KPFM (pyroelectric potentials can easily exceed tens of volts for a small change in temperature). Despite the opportunity for HV-KPFM for materials explorations, literature shows a striking dearth of attempts to make a working HV-KPFM system.

There have been some efforts, using external feedback loops, HV sources, and Arduino-based software. These endeavors have primarily focused on examining line profiles of surface potentials along the channels of organic thin-film transistors.[9] The addition of external feedback loops, however, can often give rise to electronic offsets at the AFM controller inputs/outputs and electronic cross talk and therefore requires a rigorous design- and troubleshooting- methodology to remove any offset-induced artefacts. A feedback-free HV modality (KPFM-HV) is also available on certain commercial AFM systems, which does not rely on the use of a HV source. Instead, it utilizes an oscillating AC voltage applied to the probe and calculates the electrostatic potential by measuring the cantilever response at the first- and second harmonics.[10] This process necessitates driving the cantilever significantly away from its mechanical resonance frequency, to avoid harmonic coupling and cantilever-crosstalk issues. In this case, the CPD is calculated using the amplitudes of first- and second harmonics simultaneously captured *via* two separate lock-in amplifiers thus increasing the system complexity. The lack of a fully developed High-Voltage KPFM, offered on standard AFMs, may also stem historically from clear identification of problems where such measurements could help reveal new phenomena or enable new physical insight. The conspicuous absence of HV-KPFM as a standard operational mode thus underscores a potentially significant gap in our understanding of material behavior at elevated surface potentials and raises questions about the extent to which HV-KPFM has been underutilized in scientific research.

In this work, we have extended the KPFM voltage regime on a commercial atomic force microscope to ±150 V. First, we demonstrate proof of principle experiments under known, controlled conditions: we show that our HV-KPFM technique can accurately measure the surface potential applied to a metallic pad and use the technique to track surface potential profiles in the high voltage regime, by independently measuring the potential difference across an inter-electrode gap where a large (50V) applied voltage decays to ground within a single scan window. Then, we show that our technique is applicable to real material characterisation problems, by extending recent experiments from literature to the high voltage regime. Specifically, we track the surface potential build-up and charge neutralisation in Polytetrafluoroethylene (PTFE) films when bombarded by ionic charges, and we spatially map the potential barriers across grain boundaries in positive temperature coefficient of resistivity (PTCR) ceramics. Finally, we introduce and demonstrate a novel, HV-KPFM based approach for the measurement of pyroelectric coefficients in bulk single crystals, by measuring the voltages developed on the terminating surfaces of pyroelectric materials upon heating and cooling. The results clearly demonstrate that HV-KPFM can be used as a novel and effective tool to fill in existing gaps in surface potential and potential gradient measurements, while also opening pathways for novel studies in materials physics.

**1.1 Methodology**

In KPFM, a metallic AFM tip makes two consecutive passes over the sample surface. The first is a traditional tapping mode topography measurement, where the tip is mechanically oscillated as it moves across the surface. Changes in topography are counteracted (and therefore inferred) by a feedback mechanism, which extends or retracts the tip to keep the oscillatory motion constant. In the second pass, the tip retraces the sample topography, this time at a constant, user-defined lift height, meaning the tip-sample distance remains constant, even over rough topography. An AC electrical bias is applied to the tip, with a frequency close to its mechanical resonance frequency, to induce oscillatory motion via capacitive forces between the tip and the sample. The first harmonic of this oscillatory force depends on the D.C. potential difference between the tip and sample, so by supplying an additional offset D.C. voltage to the tip ($V_{DC}$, which is equal to the tip-sample potential difference), the oscillation is nullified. In this case, the oscillation of the tip, at the frequency of the applied AC bias, acts as the feedback signal for the applied DC bias. The map of the nullifying DC bias across the surface is a direct image of the true surface potential. Importantly, the magnitude of the voltage limit of the hardware of most off-the-shelf AFMs is 10V. As KPFM can only measure what can be applied to the AFM tip, so this becomes the de facto measurement limit.

Here, we have employed a high voltage (HV) module, custom-designed for the MFP-3D Infinity system, around which the HV-KPFM experiment is designed. A customised isolated stage (Figure S1, Supp. Mater.) is used in the experiment which carries the high voltage generated using a high voltage amplifier integrated on to a relay enabled card which then communicates directly to the Asylum Research Controller (ARC). The HV tip holder (Figure S1, Supp. Mater.) is also custom designed, such that the bias supplied to the HV stage is directly routed to the metallic tip, which in turn is fully isolated from the electronics that mechanically drive the tip (thus enabling KPFM). This isolation of the high voltage to the tip is crucial in protecting the remaining circuitry of the AFM head. The HV module generates precise AC+DC signals between -150V to +150V with no significant delays meaning HVKPFM can be performed at speeds comparable to conventional KPFM. The advantage that this approach has, over alternative methods, is that the AFM controller communicates directly with both the piezo control and the high voltage module (which supplies DC voltage to the tip), making the feedback loop more efficient and reliable than introducing external electronics and feedback loops. During scanning in the interleave pass, the appropriate mix of AC and DC voltages is directly routed to the tip through the AFM software. The software control panel has been customised, such that the nullifying voltage can be directly recorded at each pixel in a precise manner. As a result, precise measurements of the surface potential could then be made possible on the nanometer scale enabling spatially resolved HV-KPFM.

## 2. Results and Discussion
### 2.1. Proof-of-concept

A simple proof-of-concept experiment demonstrating HV-KPFM is shown in Figure 1. A DC voltage, supplied by an external power supply unit, was fed to a planar gold electrode, sputtered onto a glass slide. A standard Pt-Ir coated AFM tip was then brought into contact with the electrode, and HV-KPFM was performed at a single spot on the electrode (a "point scan", as illustrated in figure 1a). The bias fed to the electrode ranged between -40V and 40V, which is accurately mapped by the measured HV-KPFM potential (fig 1b). The data points and errors in figure 1b are taken as the mean and standard deviation from 256x256 KPFM data points, all taken at the same x-y position on the electrode. Errors from the mean value are on the order of ±0.06V, which include errors from crosstalk with the topography signal.[11,12] It is suspected that this crosstalk contributes to the majority of the error.

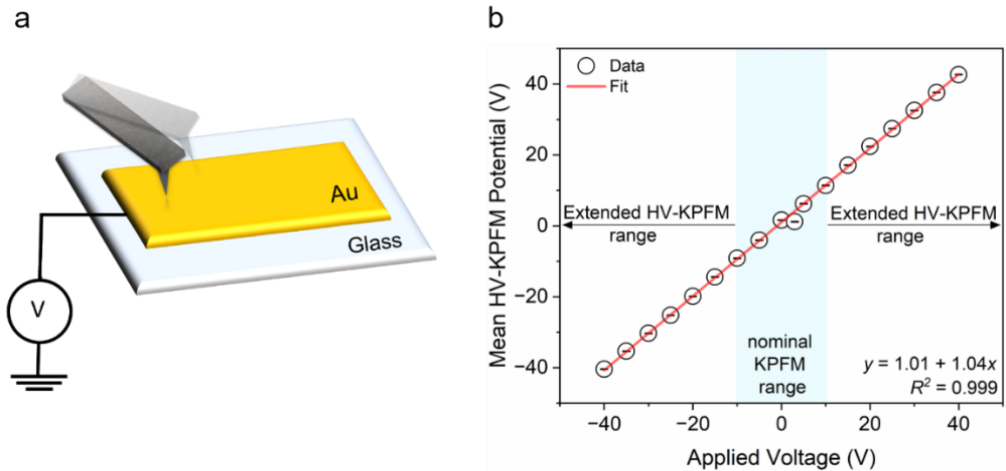

**Figure 1**: An illustration of surface potential measurement in the high voltage regime. (a) The measurement is undertaken on a single spot on the gold electrode which has bias applied via a bias source (b) Measured surface potential using HV-KPFM setup shows excellent agreement with the applied voltage.

HV-KPFM has the capability not only to measure large surface potentials, but to accurately resolve the spatial profile of large surface potentials with nanoscale resolution. To show this, a gold bar was sputtered, again onto a glass slide, and an interelectrode gap was AFM machined into the bar, using established AFM tip-based machining techniques. [13, 14] A large potential difference was then applied between the two, now separated, electrodes. The process is schematically represented in Figure 2a. Figure 2b shows a HV-KPFM surface potential map of the interelectrode region. The map shows clearly a relatively flat potential profile of ~40V (0V) measured on the high (low) electrode, and a linear potential drop across the interelectrode gap. Figure 2c plots three line sections taken across the interelectrode gap. The values match those that we might expect for the externally applied voltage of 40V. This experiment validates that HV-KPFM has the ability to map potential profiles in the high voltage with the same high spatial resolution as typically afforded by conventional atomic force microscopy techniques. In principle, this neatly extends the range of physical phenomena that can be examined with surface potential mapping by KPFM.

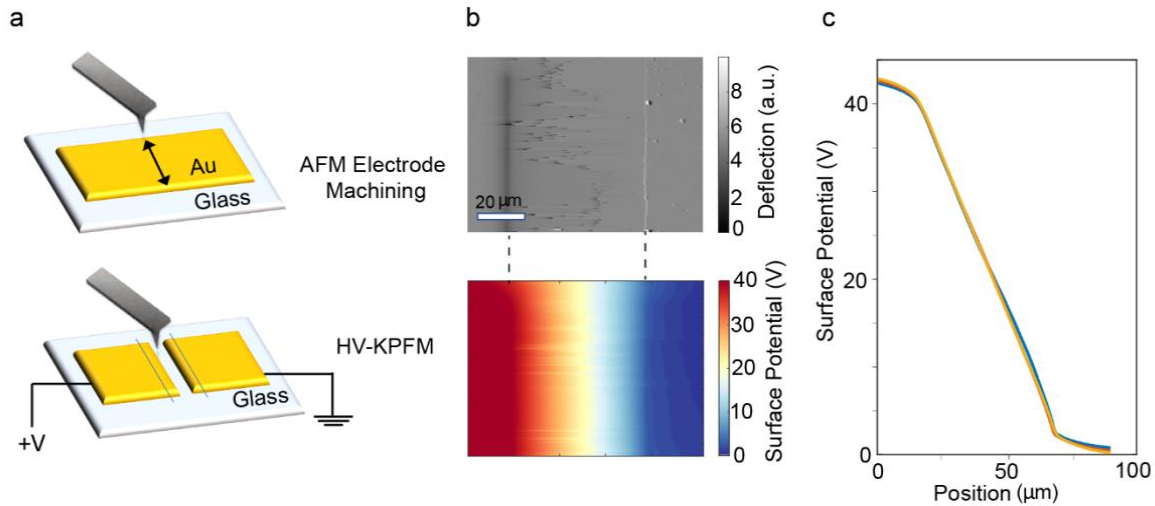

**Figure 2**: HV-KPFM based potential gradient mapping (a) Schematic illustrating AFM based milling to create an interelectrode gap when potential gradient is mapped (b) HV-KPFM potential map of the region indicated in (a) showing steep drop of the applied potential across the electrodes (c) three line profiles of the potential drop across the gap are observed to be linear in line with expectations.

## 2.2. Grain boundary potential mapping in PTCR

The ability to precisely map surface potentials in lateral geometries where bias is applied across electrodes brings out interesting possibilities in terms of exploring transport in materials. Typically, when a lateral electric field is applied to a material with homogeneous electrical properties, the change in potential is linear (as seen in Figure 2c). However, inhomogeneity in the linearity of the potential profile can be used to evaluate the role of microstructures in electrical transport[15] as well as determination of electronic transport characteristics in the presence of magnetic fields. A classic example of the latter case is a KPFM based Hall-potential detection technique, as utilised to investigate the carrier transport at charged conducting ferroelectric domain walls.[8, 16] In addition, the spatial distribution of electric potential dropped along curved current carrying conducting domain walls in classic insulators such as $LiNbO_3$ can be mapped using KPFM in the high voltage regime and such efforts can even unveil physics associated with ferroelectric domain wall p-n junctions (otherwise very difficult to characterise).[17] Moreover, HV-KPFM could in principle allow the visualisation of more extreme electronic transport regimes (such as ballistic transport), which become evident only under high electric fields, involving potential drops beyond the limits of conventional KPFM.

In prior work undertaken by our team, conventional KPFM was used to investigate the role of grain boundaries in a PTCR ceramic in a lateral geometry where they form part of the electrical circuit.[15] It was shown that, even in the low-resistance ferroelectric state, the potential drop at grain boundaries is significantly greater than in the grain interiors. Due to limitations of examinable bias range in KPFM, it was not possible to map the entire interelectrode gap in such measurements as significant field drop occurred at electrodes. Here, we have employed HV-KPFM to evaluate the entire field profile across the interelectrode gap on polished PTCR ceramic with up to 40V applied in lateral geometry. Our goal was also to accentuate the non-linear change in the field associated with the grain boundaries and how it evolves as the applied bias is increased. Figure 3a shows a topography map of a polished region on the PTCR ceramic with the corresponding grain boundaries outlined in Figure 3b. The spatial profile of the potential and its derivative are shown in Figs. 3c and 3d respectively. It is worth noting that a significant fraction of the applied bias is dropped at the right-hand electrode/ceramic junction, as evidenced by the concentration of the field in Fig. 3d. Beyond this junction, the voltage drop across the gap has a more linear trend and the potential derivative map shows enhanced drops at the grain boundaries, a result that confirms that these interfaces are indeed much more resistive than the bulk (interior) of the grain. A trimmed section of the derivative maps obtained, at increasing biases, shows that the potential drop (and hence the local field) increases at the grain boundaries. It is worth pointing out that the gradient maps look qualitatively different at higher biases with the contour lines more concentrated near the grain boundaries. Thus, HV-KPFM enables us to measure the changes in potentials at grain boundaries more vividly and further reinforce their role in mediating the conduction in PTCR ceramics (in congruence with the Heywang-Jonker model).[18]

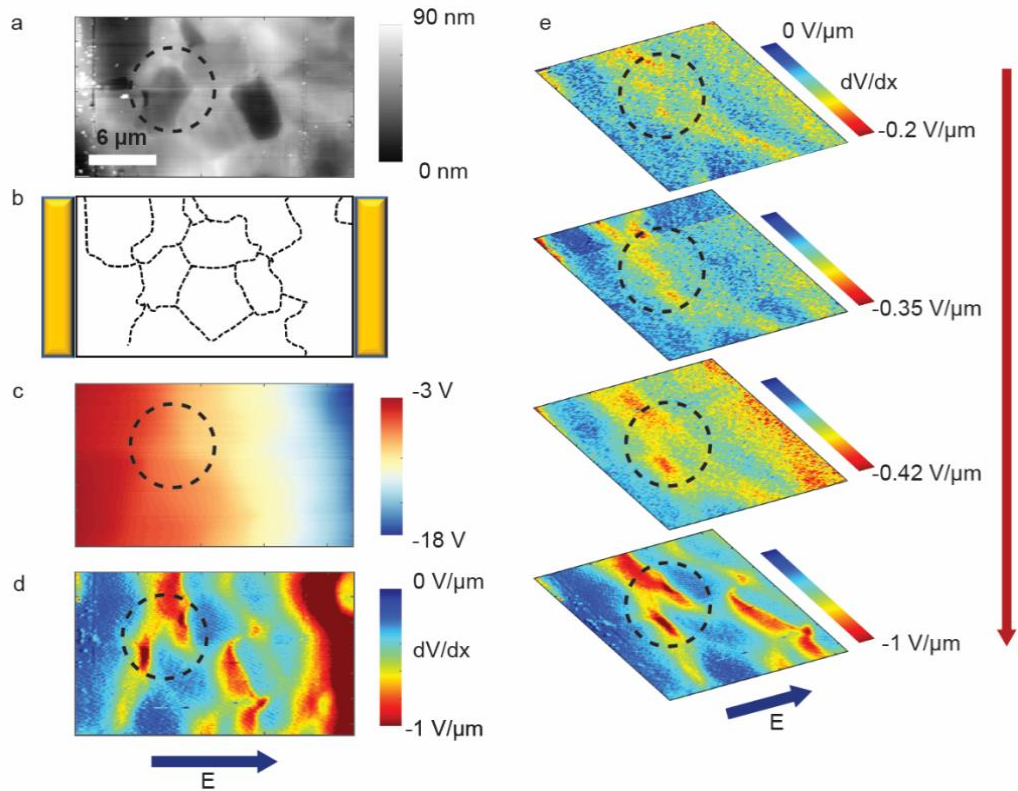

**Figure 3**: Mapping grain-boundary mediated electrical transport in PTCR ceramics. (a) Topography of the BTO-PTO-CTO surface for the KPFM experiment (26 × 13 μm), with Au electrodes placed parallel to the scan direction. (b) Schematic of the scanned area, (c) Potential map acquired by applying −20V across the surface of the sample, measured using HV-KPFM and (d) gradient of the potential map. The concentration of field lines on one electrode illustrate that a significant fraction of the applied voltage drops at the electrode-ceramic junction. (e) Gradient maps with increasing electric field mapped via HV-KPFM demonstrate the key role of grain boundaries in determining electrical transport in the PTCR ceramic.

**2.3. Assessing static charge-induced surface potential in insulators**

The HV-KPFM technique can also be extended to study charging dynamics in materials in regimes beyond ±10V. Typically, highly crystalline fluoropolymers, such as PTFE, fluorinated ethylene propylene and polyvinylidene fluoride, exhibit the highest electron affinities thus leading to remarkably large surface potentials. [19] As such, utilising electret, piezoelectric, and triboelectric phenomena, these materials form the backbone of energy harvesting platforms. [20] Particularly, tribo-negative PTFE is the material of choice for sliding and contact-mode nanogenerators, wherein nano-structuring or corona-discharge processes are undertaken to enhance its macroscopic exchange charge density. [21, 22] However, the nanoscale imaging of the induced changes is limited to either qualitative electrostatic force microscopy or the restricted conventional KPFM imaging, neither of which allows quantitative nanoscale spatial evaluation of the surface potential. We have assessed the applicability of HV-KPFM to study the surface potential values for as-obtained PTFE but to also track and quantify charge injection (and neutralisation) processes. For thin PTFE films (12.7 µm) the typical laminar arrangement of the globular structures visible in the topography (left column of Fig. 4b, upper row) is not reflected in the potential map owing to channel saturation (left column of Fig. 4b, lower row). In comparison, HV-KPFM (see right column of Fig. 4b, lower row) was able to successfully map and quantify this surface potential in the range of -34 to -40V, whose origins lie in the stretching induced orientational effects and resultant charge trapping characteristics. To replicate the effects of nano-structuring/high-voltage induced enhanced charge density, we utilise a charge injection process with controlled polarizability. Figure 4c-e shows the HV-KPFM surface potential maps of the PTFE sample at different stages of the charge injection process. First, the surface was neutralised (using positive charges) consequently lowering the surface potential to approximately -9.4V. The subsequent negative charge injection cycles significantly increased the surface potential value in the negative regime to approximately -119.9 V. Lastly, positive charges were injected, to raise the surface potential to a value of +68.7V. These charge states arose from the presence of trapped, uncompensated charges on the surface and could easily be manipulated, and monitored, as shown here. [23] The associated charge densities could be harnessed to make more efficient triboelectric nanogenerator devices with still higher voltage outputs.

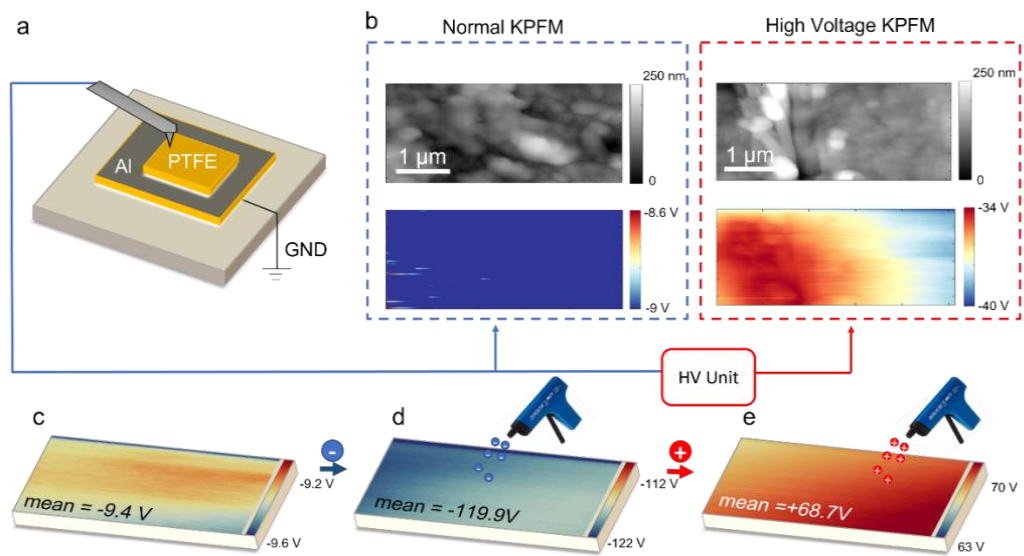

**Figure 4**: Evaluating charge injection in polymer films using HV-KPFM. (a) Schematic of spatially resolved HV-KPFM on Polytetrafluoroethylene (PTFE) film (b) Topography and conventional KPFM potential map which is saturated and unable to measure the true surface potential in this film compared with HV-KPFM data which shows correlation of surface potential with microstructure in -34 to -40V regime (c) spatially resolved

HV-KPFM potential maps upon charge injection (d) and neutralisation (e) with a charge gun illustrate the effective surface potential mapping on the nanoscale in this material.

## 2.4. HV-KPFM assessment of pyroelectric coefficient of ferroelectrics

The presented results illustrate the direct implications of HV-KPFM for evaluating material physics via quantitative measurement of surface potentials in the high voltage regime. The developed technique can also open up new ways of measuring material properties linked to surface potential, where conventional KPFM simply could not be used. In this context, we have employed HV-KPFM to measure pyroelectric coefficients of ferroelectric crystals. Pyroelectric materials exhibit a change in the spontaneous polarisation vector in response to temperature changes.[24] This results in a change in the bound surface charge on the terminating crystal surfaces normal to the polarisation. Typically, pyroelectric behaviour in ferroelectrics is measured by measuring the voltage across a reference capacitor or by measuring the pyroelectric current through a reference resistor, upon changes in temperature. Undertaking this measurement on the nanoscale requires a metallic tip to be able to accurately read the potential on the surface. Prior attempts to employ KPFM in measuring pyroelectric behaviour in ferroelectrics have been limited to thin films[25] due to the possibility of large potential build-up in single crystals for small changes in temperatures.

The 'true' pyroelectric coefficient of a material can be expressed (under constant stress) as shown in equation (1). It is worth noting that this value is different from the 'generalised' pyroelectric coefficient which relates dielectric displacement to changes in temperature.

$$p = \left(\frac{\partial P}{\partial T}\right)_{s,i} = \left(\frac{\partial P}{\partial E}\right)_{s,i} * \left(\frac{\partial E}{\partial T}\right) \qquad \text{Equation 1}$$

where $p$ refers to the true pyroelectric coefficient, P refers to spontaneous polarisation, T indicates temperature and E refers to the electric field. For a monodomain ferroelectric crystal of thickness $d$, this expression further reduces (with suitable assumptions) to that shown in equation (2).

$$p = \varepsilon_0 \varepsilon_r \frac{\Delta V}{d * \Delta T} \qquad \text{Equation 2}$$

Thus, the change in surface potential $\Delta V$ arising from a temperature change of $\Delta T$ is given as follows:

$$\Delta V = \frac{p\,\Delta T\,d}{\varepsilon_0 \varepsilon_r} \qquad \text{Equation 3}$$

where $\varepsilon_r$ is the dielectric constant and $\varepsilon_0$ indicates the permittivity of vacuum. An important assumption made to arrive at equation 3 is that the chosen permittivity value assumes a linear dielectric behaviour regime for the examined sample. Reported pyroelectric coefficients for single crystal LiNbO$_3$ (along the polar [001] axis) lie between -41 μC/(m$^2$-K) and -83 μC/(m$^2$-K). [26-29] This means that for a 175μm thick z-cut crystal, a potential difference of 19 to 37V should be expected for just 1K change in temperature on the [001] face, rendering conventional KPFM ineffective (as the surface potential scales with crystal thickness, potentials of up to 3kV can be observed for a 10mm crystal of LNO for just 1K change in temperature). Hence, HV-KPFM offers an ideal nanoscale platform to measure pyroelectric coefficients in ferroelectric crystals, and furthermore, offers the functionality to measure spatial variations in pyroelectric properties.

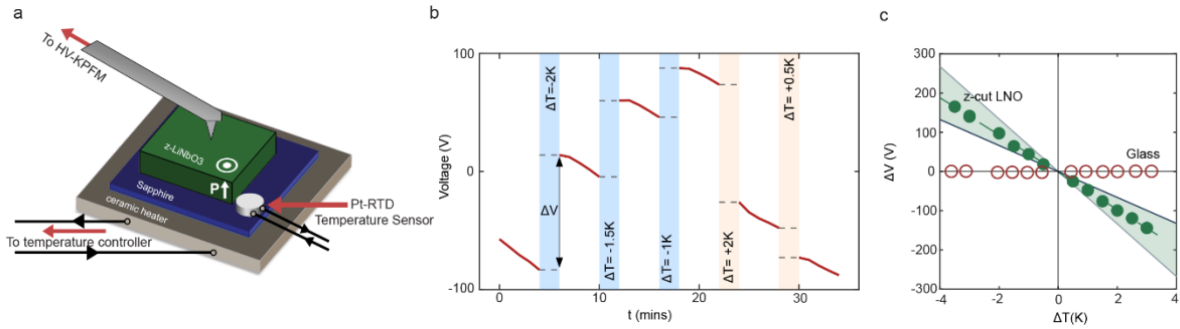

**Figure 5**: Measurement of pyroelectric coefficient in a ferroelectric single crystal of LiNbO$_3$ via HV-KPFM. (a) Schematic illustrating a z-cut LiNbO$_3$ crystal on a temperature controlled ceramic heater with HV-KPFM used to measure changes in surface potential quantitatively after each temperature step ($\Delta$T) (b) The change in voltage $\Delta$V after each temperature step is measured (c) Plot of $\Delta$V vs $\Delta$T gives pyroelectric coefficient of -63 μC/m$^2$-K which lies between range of values reported for LiNbO$_3$ crystals (-41 to -83 μC/m$^2$-K). [26-29] A reference sample of glass shows no pyroelectric behaviour in line with expectations.

The setup for measurement of pyroelectric coefficients in a ferroelectric single crystal of LiNbO$_3$ via HV-KPFM is illustrated in Figure 5a. A z-cut monodomain LiNbO$_3$ crystal is polished to a thickness of 175μm and placed on a sapphire crystal which in turn sits on a ceramic heater, whose temperature can be accurately controlled (to within ~0.1K) and measured via a temperature sensor. With each temperature step, the change in surface potential is measured. As shown in Figure 5b and c, change in voltage is linear with the change in temperature and reveals a pyroelectric coefficient of -59 μC/(m$^2$K). This observed value sits in the middle of reported values for pyroelectric coefficient of LiNbO$_3$ indicated by the shaded region shown in Figure 5c. Similar studies were undertaken on LiTaO$_3$ crystals (higher pyroelectric coefficients, see Table S1 in Supp. Mater.) as well as amorphous glass substrates (no response, see Table S2 in Supp. Mater.) and results were found to be consistent with expectations (see section 2 in Supp. Mater.). In essence, HV-KPFM offers a unique approach to measure pyroelectric coefficients in single crystal ferroelectrics and could, in principle, enable nanoscale spatially resolved mapping of pyroelectric behaviour in ferroelectric crystals with complex microstructures.

## 3. Conclusions

In conclusion, we have introduced and investigated a high-voltage Kelvin probe force microscopy technique, capable of reliably measuring surface potentials well beyond the conventional 10V limit. We have demonstrated the accuracy of the technique, by mapping known potential profiles across interelectrode gaps, in excess of 10V. We have also applied HV-KPFM to real material characterisation problems; we map the charging and discharging in PTFE (a material key for triboelectric voltage generation), and we image the functional response of grain boundaries in positive temperature coefficient of resistivity (PTCR), all in a voltage regime inaccessible to regular KPFM. Finally, we proposed a novel method for the measurement (and spatial mapping) of pyroelectric coefficients in ferroelectric single crystals, by using HV-KPFM to measure the generated pyroelectric voltage on heating and cooling. We believe the extension of surface potential measurement by KPFM to the high voltage regime will greatly expand the range of material processes accessible by KPFM, unlocking a new world of spatially resolved high voltage surface phenomena.

## 4. Experimental Methods

*Sample preparation*: The sample used in PTCR mapping composed of BaTiO$_3$–PbTiO$_3$–CaTiO$_3$ (BTO-PTO-CTO) where the percentage concentrations are 68%, 20%, and 12% respectively, with an average measured grain size diameter of 4 $\mu$m. The ceramics typically have relatively low bulk resistivity at room temperature (10–100 Ω cm). Samples for this study were prepared from bulk by cutting and polishing using diamond paper and a colloidal silica solution. Gold electrodes were deposited with an interelectrode gap of ~40 µm via sputtering and KPFM was undertaken in the gap. For the pyroelectric measurements, a z-cut lithium niobate single crystal (obtained from MTI corporation, USA) was polished to the desired thickness.

*Tip-based electrode milling*: For the investigations in Figure 2, an interelectrode gap was machined into a deposited gold electrode, by AFM micromachining. A silicon AFM tip with a cantilever of stiffness constant of 40N m$^{-1}$ was scanned over the metallic bar in contact mode. A large deflection set-point was used, which resulted in an enhanced compressive force (of ~ 1 µN) being imparted from the tip to the sample. This results in the line-by-line removal the metallic film as the AFM tip scans. The scan window was positioned to generate an approximately 50 µm gap in the deposited metallic bar.

*Kelvin Probe Force Microscopy*: For conventional KPFM studies, an Asylum Research MFP-Infinity Atomic Force Microscope was used in the amplitude-modulated KPFM mode to perform spatially resolved surface potential mapping in two-pass mode. For such measurements, a Pt-coated Si tip (Nanosensor PPP-EFM) with a stiffness constant k of 2.8 N m$^{-1}$ was used (typical resonance frequency around 70 kHz). The HV module was used with the same tips for the high voltage KPFM measurements.

*Heating the pyroelectric crystal*: For the pyroelectric measurements, the sample is mounted on a sapphire substrate, which in turn is then mounted on a temperature-controlled ceramic heater. A platinum resistance temperature detector (Pt100-RTD) is used to allow the temperature to be measured and accurately controlled within 0.1K via a Thor Labs TC200 temperature controller.

**Acknowledgements**: A.K. and K.H. gratefully acknowledge support from Department of Education and Learning NI through grant USI-205 and Engineering and Physical Sciences Research Council via grant EP/S037179/1. J.M.G is grateful for the financial support received from the Engineering and Physical Sciences Research Council (EPSRC through grant EP/P02453X/1 and through studentship funding). R.McQ gratefully acknowledges support via UKRI Future Leaders Fellowship program (MR/T043172/1). The authors gratefully acknowledge the financial support received from Tezpur University, India in the form of PhD studentship to NS under the collaborative TU-QUB PhD program. B.J.R. gratefully acknowledges support from Science Foundation Ireland via SFI/21/UUS/3765.

**Data availability**: The data that supports the findings of this study are available from the corresponding author upon reasonable request.

**Spatially Resolved High Voltage Kelvin Probe Force Microscopy: A Novel Avenue for Examining Electrical Phenomena at Nanoscale**

*Supplementary Information for the main manuscript*

*Conor J. McCluskey[#], Niyorjyoti Sharma[#], Jesi R. Maguire, Serene Pauly, Andrew Rogers, Tj Lindsay, Kristina M. Holsgrove, Brian J. Rodriguez, Navneet Soin, John Marty Gregg, Raymond G. P. McQuaid, and Amit Kumar\**

E-mail: a.kumar@qub.ac.uk
[#] These authors contributed equally to the work

C. McCluskey, N. Sharma, J. Maguire, S. Pauly, A. Rogers, Tj Lindsay, K. M. Holsgrove, J. M. Gregg, R. G. P. McQuaid, A. Kumar
School of Mathematics and Physics, Centre for Quantum Materials and Technologies, Queen's University Belfast, Belfast, BT7 1NN, United Kingdom

N. Soin
School of Engineering, Ulster University, 2-24 York Street, Belfast BT15 1AP, United Kingdom
School of Science, Computing and Engineering Technologies, Swinburne University of Technology, P.O. Box 218, Hawthorn VIC 3122, Australia

B. J. Rodriguez
School of Physics, University College Dublin, Belfield, Dublin 4, Ireland

This document contains additional details on the implementation of High Voltage Kelvin probe force microscopy (HV-KPFM) and the determination of pyroelectric coefficients on ferroelectric crystals using the technique.

# 1. High Voltage Kelvin Probe Microscopy setup (on Asylum Research MFP-3D Infinity system)

We have employed a customised high-voltage (HV) source which communicates directly with the Asylum Research controller for the MFP-3D Infinity system, without the need to incorporate other external triggers or additional amplifiers. The HV source delivers the applied potential to the stage which in turn is connected to the tip. It is worth noting that the rest of the electronics on the tip holder are partitioned from the HV circuit thus enabling smooth operation of the microscope in the HV-KPFM mode. In essence, this creates a closed-loop version of the KPFM and allows systematic and robust surface potential mapping in the high-voltage regime. The components enabling the implementation of HV-KPFM are shown in Figure S1 below.

The software interface enabling electrical routing connections (called cross-point panel in MFP-3D infinity) through the HV source is customised to send appropriate signals to the tip during the second pass of the KPFM measurement.

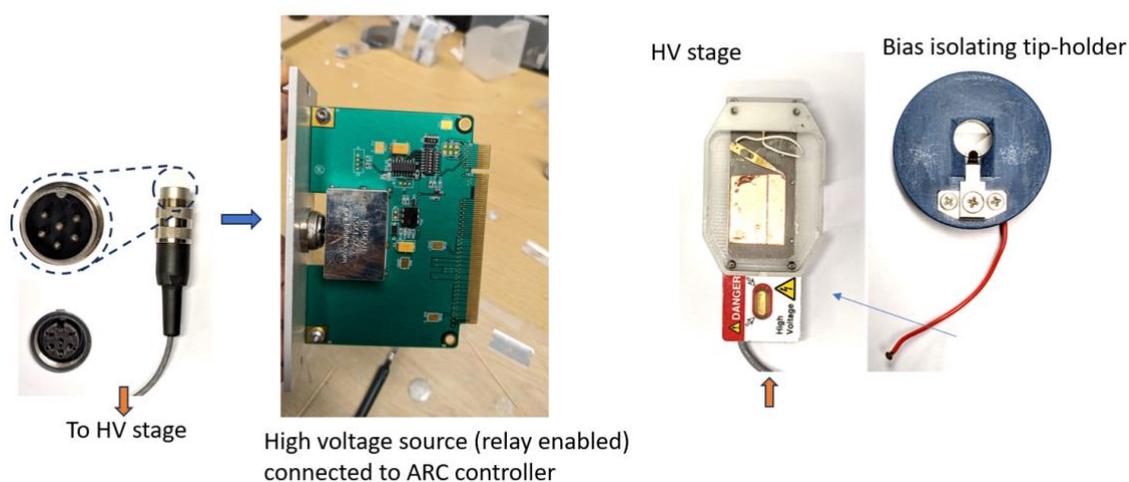

*Figure S1: The configuration enabling closed-loop HV-KPFM on Asylum Research MFP-3D Infinity system.*

# 2. Employing HV-KPFM to measure pyroelectric coefficients in LiTaO$_3$ crystals

In the main manuscript, we have demonstrated the measurement of the pyroelectric coefficient of a LiNbO$_3$ crystal using HV-KPFM. This experiment was repeated on a single crystal *z*-cut LiTaO$_3$, with the results reported in *Table 1* and *Figure 5*. The expected pyroelectric coefficient for LiTaO$_3$ is -176μCm$^{-2}$K$^{-1}$, with the experimental results giving a value of -150μCm$^{-2}$K$^{-1}$.

**Table S1:** HV-KPFM measured ΔV (*vs*. ΔT) values for a single crystal *z*-cut LiTaO$_3$

*LiTaO$_3$ data*

| ΔT (K) | ΔV (V) |
|---:|---:|
| -2 | 133 |
| -1.5 | 103.4503 |
| -1 | 69.4357 |
| -0.5 | 35.619 |
| 0.5 | -32.7876 |
| 1 | -58.2187 |
| 1.5 | -95.8396 |
| 2 | -126.37 |

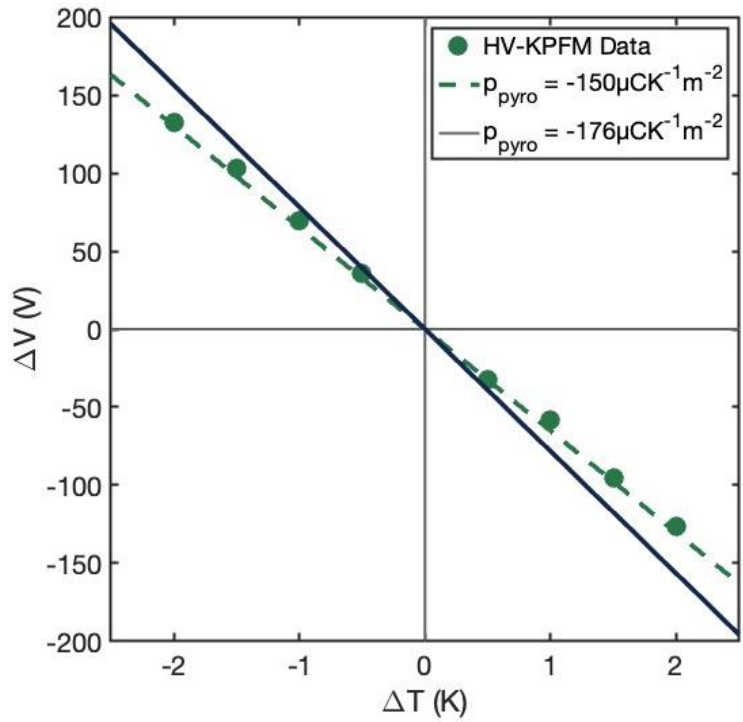

*Figure S2: Plot of ΔV vs ΔT for a single crystal z-cut LiTaO$_3$ gives the pyroelectric coefficient of -150 µC/m$^{-2}$K$^{-1}$. The literature value is -176 µC/m$^{-2}$K$^{-1}$.*

To ensure that these potentials are developed because of the pyroelectric effect and not some experimental quirk, a small piece of glass was used as a test sample. The results are seen in *Table 2*, and they confirm that the potentials developed are not a result of any experimental features, but rather must be attributed to the pyroelectric effect. The minimal shifts in voltage seen in the glass results are attributed to small expansion or contraction in the glass.

**Table S2:** HV-KPFM measured ΔV (*vs.* ΔT) values for a glass substrate

| Glass data ||
|---:|---:|
| ΔT (K) | ΔV (V) |
| 1 | 0.69 |
| 2 | 0.8442 |
| -1 | -0.6637 |
| -2 | -0.7765 |